\documentstyle[prl,aps]{revtex}
\begin{document}
\draft
\preprint{gr-qc/9911084}
\title{Non--Singular Charged Black Hole Solution for Non--Linear Source}
\author{Eloy Ay\'on--Beato and Alberto Garc\'{\i}a}
\address{Departamento~de~F\'{\i}sica,~Centro~de~Investigaci\'on~%
y~Estudios~Avanzados~del~IPN\\
Apdo. Postal 14--740, 07000 M\'exico DF, MEXICO}
\maketitle

\begin{abstract}
A non--singular exact black hole solution in General Relativity is
presented. The source is a non--linear electromagnetic field, which reduces
to the Maxwell theory for weak field. The solution corresponds to a charged
black hole with $|q|\leq 2s_{{\rm {c}}}m\approx 0.6\,m$, having metric,
curvature invariants, and electric field bounded everywhere.
\end{abstract}


The existence of singularities is one of the basic failure of the General
Relativity Theory \cite{H-E}; it appears to be an inherent property to most
of the solutions of the Einstein equations. The Penrose censorship
conjecture establishes that these singularities must be dressed by event
horizons; no causal connection exits with the interior of a black hole, and
thus the pathologies occurring at the singular region have no influence on
the exterior region, and the Physics outside is well--behaved 
(cf.~\cite{Wald97} for a review on the recent status of this conjecture).
Nevertheless, the whole space--time has to contain also the interior of
black hole, since gravity permits physical objects to fall inside. Hence, we
need to know what happens in this falling process. However, the singular
behavior of the known black hole solutions made impossible a good
description of it, this has been interpreted has a breakdown of General
Relativity.

Some regular black hole models has been proposed \cite
{Bardeen68,Ayon93,Borde94,BarrFrolov96,MMPSenovilla96,CaboAyon97} in order
to understand this process. All of them have been referred as ``Bardeen
black holes'' \cite{Borde97}, since Bardeen was the first author to produce
a surprisingly regular black hole model \cite{Bardeen68}. No one of these
models is an exact solution to Einstein equations; there is no known
physical sources associated with any of them. The solution to this problem
has usually been suggested by finding more general gravity theories avoiding
the existence of singularities. The best candidate today to produce
singularity--free models, even at the classical level, due to its intrinsic
non--locality is string theory \cite{Tseytlin95}. There are examples in
other contexts, for instance, in $N=1$ supergravity domain wall solutions
with horizons but no singularities have been found (cg.~\cite{Cvetic93}
and references therein), other example is in exact conformal field theory 
\cite{HHorowitz92}. 

We show in this letter that there is no need to desist from General
Relativity to solve the singularity problem. By assuming an appropriate
non--linear source of matter, which in the weak field approximation becomes
the usual linear theory, one can achieve a singularity--free black hole
solution to Einstein equations coupled with a non--linear electrodynamics.
Previous attempts in this direction with non--linear electrodynamics either
have totally been unsuccessful or only partially solve the singularity
problem \cite{Oliveira94,Soleng95,Palatnik97}.

We derived our solution using a non--linear electrodynamic source described
by the action \cite{SGP87} 
\begin{equation}
{\cal S}=\int dv\left( \frac 1{16\pi }R-\frac 1{4\pi }(2P{\cal H}_P-{\cal H}%
)\right) ,  \label{eq:action}
\end{equation}
where $R$ is scalar curvature, $P\equiv \frac 14P_{\mu \nu }P^{\mu \nu }$, 
\begin{equation}
{\cal H}(P)=\frac{P\,{\rm {e}}^{-s\sqrt[4]{-2q^2P}}}{\left( 1+\sqrt{-2q^2P}%
\right) ^{5/2}}\left( 1+\sqrt{-2q^2P}+\frac 3s\sqrt[4]{-2q^2P}\right) 
\label{eq:H}
\end{equation}
is a function describing the source, and ${\cal H}_P\equiv \partial {\cal H}%
/\partial P$. In (\ref{eq:H}) $s\equiv |q|/2m$, $q$ and $m$ are free
parameters which will be associated with charge and mass respectively. The
last function satisfies the plausible condition, needed for a non--linear
electromagnetic model, of correspondence to Maxwell theory, i.e., $%
{\cal H}\approx P$ for weak fields ($P\ll 1$). In this description the usual
electromagnetic strength is given by $F_{\mu \nu }\equiv {\cal H}_PP_{\mu
\nu }$.

In order to obtain the desired solution we consider a static and spherically
symmetric configuration 
\begin{equation}
\mbox{\boldmath$g$}=-A(r)\mbox{\boldmath$dt$}^2+A(r)^{-1}\mbox{\boldmath$dr$}%
^2+r^2\mbox{\boldmath$d\Omega $}^2,  \label{eq:spher}
\end{equation}
and the following ansatz for the anti--symmetric field $P_{\mu \nu }=2\delta
_{[\mu }^0\delta _{\nu ]}^1D(r)$. With these choices the
Einstein--non--linear electrodynamic field equations following from action (%
\ref{eq:action}), 
\begin{equation}
G_\mu ^{~\nu }=2({\cal H}_PP_{\mu \lambda }P^{\nu \lambda }-\delta _\mu
^{~\nu }(2P{\cal H}_P-{\cal H})),\qquad \nabla _\mu P^{\alpha \mu }=0,
\label{eq:Eqs}
\end{equation}
are directly integrated, yielding 
\begin{equation}
\mbox{\boldmath$g$}=-\left( 1-\frac{2mr^2{\rm {e}}^{-q^2/2mr}}{%
(r^2+q^2)^{3/2}}\right) \mbox{\boldmath$dt$}^2+\left( 1-\frac{2mr^2{\rm {e}}%
^{-q^2/2mr}}{(r^2+q^2)^{3/2}}\right) ^{-1}\mbox{\boldmath$dr$}^2+r^2%
\mbox{\boldmath$d\Omega $}^2,  \label{eq:regbh}
\end{equation}
\begin{equation}
D=\frac q{r^2}.  \label{eq:dielec}
\end{equation}
We can note that $q$ actually plays the role of the electric charge; a
calculation of the electric field $E=F_{01}={\cal H}_PD$ gives 
\begin{equation}
E=\frac{q\,{\rm {e}}^{-\,q^2/2mr}}{(r^2+q^2)^{7/2}}\left( r^5+\frac{%
(60\,m^2-q^2)r^4}{8\,m}+\frac{q^2r^3}2-\frac{q^4r^2}{4m}-\frac{q^4r}2-\frac{%
q^6}{8m}\right) ,  \label{eq:E}
\end{equation}
from which two facts follow: the electric field is bounded everywhere, and
asymptotically behaves as $E=q/r^2+O(1/r^3)$, i.e., a Coulomb field
with electric charge $q$. With regard to the metric, it can be noted that it
asymptotically behaves as the Reissner--Nordstr\"om solution, i.e., $%
g_{00}=1-2m/r+q^2/r^2+O(1/r^3)$, so the parameters $m$ and $q$ can be
correctly associated with mass and charge respectively.

We will show that for a certain range of mass and charge our solution is a
black hole, which moreover is non--singular everywhere. Making the
substitution $x=r/|q|$, $s=|q|/2m$ we write 
\begin{equation}
-g_{00}=A(x,s)\equiv 1-\frac 1s\frac{x^2{\rm {e}}^{-s/x}}{(1+x^2)^{3/2}}.
\label{eq:A}
\end{equation}
For any value of $s$, the last function has a single minimum for $x_{{\rm {m}%
}}(s)=(s+(s^2+6)/R(s)+R(s))/3$, where $R(s)\equiv (s^3+(45/2)s+3\sqrt{%
3(s^4+(59/4)s-8)})^{1/3}$. For $s<s_{{\rm {c}}}$ this minimum is negative,
for $s=s_{{\rm {c}}}$ the minimum vanishes and for $s>s_{{\rm {c}}}$ the
minimum is positive, where $s_{{\rm {c}}}\approx 0.3$ is the solution to the
equation $A(x_{{\rm {m}}}(s),s)=0$. Calculating the curvature invariants $R$%
, $R_{\mu \nu }R^{\mu \nu }$, and $R_{\mu \nu \alpha \beta }R^{\mu \nu
\alpha \beta }$ for metric (\ref{eq:regbh}) one establishes that all of them
are bounded everywhere; thus for $s\leq s_{{\rm {c}}}$ the singularities
appearing in (\ref{eq:regbh}) (the vanishing of $A$) are only
coordinates--singularities describing the existence of horizons, and we are
in presence of black hole solutions for $|q|\leq 2s_{{\rm {c}}}m\approx
0.6\,m$. For these values of mass and charge we have, under the strict
inequality $|q|<2s_{{\rm {c}}}m$, inner and event horizons for the Killing
field $\text{\boldmath$k$}=\text{\boldmath$\partial /\partial {t}$}$,
defined by $-k_\mu k^\mu =A(r)=0$. For the equality $|q|=2s_{{\rm {c}}}m$,
they shrink into a single horizon, where also $\nabla _\nu (k_\mu k^\mu )=0$%
, i.e., this case corresponds to an extreme black hole as in the
Reissner--Nordstr\"om solution. The extension of the metric beyond the
horizons, up to $r=0$, becomes apparent by passing to the standard advanced
and retarded Eddington--Finkelstein coordinates, in terms of which the
metric is well--behaved everywhere, even in the extreme case. The maximal
extension of this metric can be achieved by following the main lines
presented in \cite{Chandra83} for the Reissner--Nordstr\"om solution, taking
care, of course, of the more involved integration of the tortoise coordinate 
$r^{*}\equiv \int {A^{-1}dr}$ in our case. Summarizing, our space--time
possesses the same global structure as the Reissner--Nordstr\"om black hole
except that the singularity, at $r=0$, of this last solution has been
smoothed out and $r=0$ is now simply the origin of the spherical
coordinates. This kind of metrics is not new (cf.~\cite{Borde97} for
a review) but the new feature in this case is that it is an {\em exact
solution\/}, as opposed to the previous ones that are only non--singular black
hole {\em models\/}.

\acknowledgments
This work was partially supported by the CONACyT Grant 3692P--E9607, and a
fellowship from the Sistema Nacional de Investigadores (SNI).

\end{document}